\documentclass[fleqn,12pt,twoside]{article}
\usepackage{espcrc1}

\usepackage{graphicx}
\usepackage[figuresright]{rotating}

\newcommand{\AmS}{{\protect\the\textfont2
  A\kern-.1667em\lower.5ex\hbox{M}\kern-.125emS}}

\title{The 3D-structure of the LISM}

\author{B. E. Wood\address[JILA]{JILA, University of Colorado and
        National Institute of Standards and Technology,
        Boulder, CO 80309-0440},
        S. Redfield\addressmark[JILA], and
        J. L. Linsky\addressmark[JILA]}
       
\begin{document}

\maketitle

\begin{abstract}

     We review what is currently known about the structure of interstellar
material in the solar neighborhood, emphasizing how observations from
the {\em Hubble Space Telescope} (HST) have improved our understanding of
how interstellar gas is distributed near the Sun.  The nearby ISM is not
uniform but shows variations in both temperature and metal abundances on
distance scales of just a few parsecs.  The observations also show
that nearby gas does not have a single uniform velocity vector.  Instead,
different components are often seen in different directions for even very
short lines of sight.  However, interpretation of these components remains
difficult.  It is uncertain whether the components represent physically
distinct clouds or perhaps are just symptomatic of velocity gradients within
the cloud.  Finally, since it is the local interstellar medium's interaction
with the solar wind that is the primary application of ISM studies
considered in these proceedings, we also review how the same HST data used
to study the local ISM structure has also been used to study both the
heliospheric interaction with the solar wind and also ``astrospheric''
interactions with the winds of other stars.

\end{abstract}

\section{INTRODUCTION}

     The Sun resides within a region of space called the Local Bubble,
which extends about 100~pc in most directions \cite{dms99}.  The existence
of the Local Bubble was first apparent on the basis of just how little
ISM material was generally observed within roughly 100~pc.  Later, the hot
material within the Bubble was detected directly from X-ray observations.
Most locations within the the Local Bubble are very hot ($T\sim 10^{6}$~K)
and rarified ($n_{e}\sim 10^{-3}$~cm$^{-3}$).  The ISM is therefore
completely ionized.

     However, the local interstellar medium (LISM)
material immediately around the Sun appears to be of a different
character.  Observations of solar Ly$\alpha$ emission scattering off
interstellar H~I gas flowing into the heliosphere revealed that the
LISM is at least partially neutral.  Furthermore, it is cooler
($T\sim 10^{4}$~K) and denser ($n_{e}\sim 10^{-1}$~cm$^{-3}$) than the
hot plasma filling most of the Local Bubble
\cite{hjf74,dpc87}.  Further studies, both of the solar backscatter
variety and of interstellar absorption lines, have further refined our
understanding of the LISM, which we consider for our purposes here to be
the warm, partially ionized ISM within roughly 10--20 pc of the Sun.
We now discuss how recent observations, especially from the {\em Hubble
Space Telescope} (HST), have continued to both clarify and complicate
our understanding of the LISM.

\section{THE ANALYSIS OF ABSORPTION LINE DATA}

\subsection{The H~I and D~I Ly$\alpha$ Lines}

     Since its launch in 1990, HST has been a
very useful source of information about the LISM, in particular the high
resolution UV spectrometers carried by the satellite, first the Goddard
High Resolution Spectrograph (GHRS) followed by the Space Telescope Imaging
Spectrograph (STIS) that replaced GHRS in 1997.  Earlier UV-based satellites,
particularly the {\em International Ultraviolet Explorer} (IUE) and
{\em Copernicus} missions provided the first extensive absorption line
studies of very nearby ISM material \cite{akd77,wbl84,wbl86,jm87,jm90,prm92}.
However, GHRS and STIS are the first UV instruments to offer spectral
resolution sufficient to {\em fully} resolve narrow LISM absorption lines,
enabling the separation of closely spaced velocity components and allowing
more accurate measurements of the absorption features.

     Because there are very few hot stars within 10--20 pc of the Sun, most
absorption line studies of the LISM rely on observations of cool stars.  The
disadvantage of working with cool star UV spectra in an ISM analysis is that
cool stars do not have strong continua in the far-UV (FUV) below 2000~\AA.
This limits the number of detectable lines, since one can only
detect an ISM line if it happens to lie on top of a stellar emission line.
The most important LISM absorption line accessible to HST is the H~I
Ly$\alpha$ line.  Analysis of H~I absorption in this spectral feature
provides column density measurements of the most abundant atom in the
LISM, and the universe in general, neutral hydrogen.  Fortunately, cool star
chromospheres provide copious amounts of H~I Ly$\alpha$ emission to provide
a background against which the ISM absorption can be seen.

     In Figure~1, we show an example of the type of HST cool star Ly$\alpha$
spectra typically used to study H~I in the LISM \cite{jll96}.  The
observations are for the two primary members of the closest star system to
the Sun, the $\alpha$~Cen system, which is only 1.3~pc away and also contains
a third distant member, Proxima~Cen.  The prevalence of H~I means that even
for the shortest lines of sight the LISM H~I absorption is always seen to be
very broad and saturated.
\begin{figure}
\includegraphics[scale=0.5,trim=150 60 0 0]{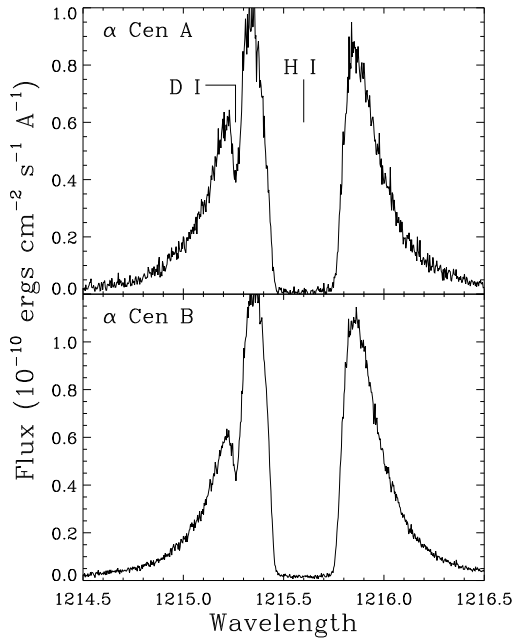}
\caption{HST/GHRS spectra of the Ly$\alpha$ lines of $\alpha$~Cen A and B,
  showing broad absorption from interstellar H~I and narrow absorption from
  D~I \cite{jll96}.}
\end{figure}

     Measurements of saturated lines are unfortunately not as precise as
measurements of weaker, unsaturated lines.  However, the opacity of H~I
Ly$\alpha$ is so high that at typical LISM column densities of
$\log {\rm N(H~I)}\approx 18.0$ (in cm$^{-2}$) the line is starting to leave
the flat part of the curve of growth.  In other words, the Lorentzian wings
of the opacity profile are starting to come into play, subtly altering the
shape of the absorption profile and allowing the H~I measurement to be
somewhat more accurate than a measurement of a line fully in the flat part of
the curve of growth.  Furthermore, the H~I analysis can also benefit if it is
combined with the analysis of the nearby deuterium (D~I) Ly$\alpha$ line,
which is the narrow, unsaturated line visible just blueward of the H~I line
in Figure~1.

     There are three free parameters in any absorption line fit:  the
centroid velocity, $v$, the Doppler parameter, $b$, and the column density,
$N$.  The Doppler parameter is connected with the temperature ($T$) and
turbulent (i.e., nonthermal) velocity ($\xi$) of the absorbing gas by the
equation
\begin{equation}
b^2=0.0165 T/A + \xi^2,
\end{equation}
where $A$ is the atomic weight of the atom or ion responsible for the
absorption, and $b$ and $\xi$ both have units of km~s$^{-1}$.  The
centroid is generally easy enough to measure for any absorption line since
it is independent of the other parameters, but for saturated lines
the Doppler parameter and column density are correlated.  This means that,
relative to a ``best fit,'' there may be fits of nearly equal quality with
lower $b$ and higher $N$, or higher $b$ and lower $N$.  Therein lies the
difficulty with accurately analyzing H~I Ly$\alpha$ by itself.

     However, the D~I line, which is generally optically thin in the LISM,
does not have this problem.  For an optically thin line, $b$ and $N$ are
essentially independent, with $b$ being associated with the width of the
absorption line and $N$ with its depth.  In section 2.2, we will describe how
comparing the $b$ values of lightweight atoms (with low $A$) with those of
heavier atoms can yield a unique measurement of $T$ and $\xi$ using equation
(1).  Such analyses clearly show that in the LISM, H~I and D~I are dominated
by thermal broadening.  This means that if the H~I and D~I Ly$\alpha$ lines
are fitted simultaneously, one can add the constraint that
$b({\rm D~I})=b({\rm H~I})/\sqrt{2}$.  This additional constraint removes
some of the dependence of $b({\rm H~I})$ on $N({\rm H~I})$, allowing both
quantities to be measured more accurately.  In such H~I and D~I fits it is
also typically assumed that the centroid velocities of H~I and D~I are the
same, which is an important constraint if the H~I absorption is affected by
a source of absorption not observed in D~I.  This is the case when
heliospheric or ``astrospheric'' material contributes to the H~I absorption
(see section 4).

     The D~I line is, of course, of interest in its own right, not just as an
aid to measuring H~I.  The deuterium-to-hydrogen ratio (D/H) in the universe
is a very important quantity both for cosmology and Galactic chemical
evolution studies \cite{tpw91,gs92,evf94,sb01}.  Thus, many of the LISM
analyses of Ly$\alpha$ lines have focused on measuring D/H, including
many of the first analyses of HST data
\cite{akd77,jm87,prm92,jll93,jll95,ard97,np97}.  One important issue concerns
how well mixed the Galactic ISM is and if there are large variations of D/H
within the ISM.  Measurements within what we are calling the LISM (i.e.,
within 10-20~pc) suggest ${\rm D/H}=(1.5\pm 0.1)\times 10^{-5}$ with little
if any evidence for variation \cite{jll98}.  This uniformity of at least the
very local ISM has allowed some Ly$\alpha$ studies of the LISM to
constrain $N({\rm H~I})$ by assuming the above D/H ratio
\cite{pb95,bew98,bew00a}, which is a reasonable thing to do if the quality of
the data is such that an accurate $N({\rm H~I})$ value cannot be derived
without constraining it with $N({\rm D~I})$.  Note, however, that there
{\em is} evidence for D/H variations on distance scales of $\sim 100$~pc and
larger \cite{avm98,ebj99,gs00}, making this assumption questionable for
longer lines of sight.

\subsection{The Metal Lines}

     In order to obtain the most accurate analysis of the Ly$\alpha$ line
it is necessary to know the number of interstellar components along the
line of sight.  Unfortunately, the large widths of both the H~I and D~I
lines (due to their low $A$ values) means these lines are not very useful
for looking for multiple ISM components along the line of sight,
since closely spaced velocity components will be too highly blended within
the H~I and D~I lines to be distinguishable.  Thus, the most accurate LISM
studies also require high resolution observations of narrow absorption lines
of heavy atoms, in addition to the Ly$\alpha$ data.

     The three most commonly observed metal lines that are useful for these
purposes are the Mg~II h \& k lines with vacuum wavelengths of 2803.531~\AA\
and 2796.352~\AA, respectively, and the Fe~II $\lambda$2600.173 line.
Figure~2 shows the Fe~II absorption observed towards $\alpha$~Cen, plotted
on a heliocentric velocity scale \cite{jll96}.
The Fe~II line was observed by HST/GHRS on two separate
occasions so there are two spectra of the same line shown in the figure.  An
absorption line fit is also shown, where the dotted line is the fit before
convolution with the instrumental profile and the thick solid line that fits
the data is the fit after the instrumental broadening is taken into account.
The difference between the two indicates the degree to which the HST/GHRS
instrument is resolving the line.  These fits require the stellar background
flux be estimated (i.e., thin solid lines in Fig.~2), but this is a
relatively easy thing to do for narrow lines like Fe~II and Mg~II.
Polynomial fits to the continuum, excluding the LISM absorption, adequately
estimate the stellar flux above the absorption.
\begin{figure}
\includegraphics[scale=0.5,trim=150 60 0 0]{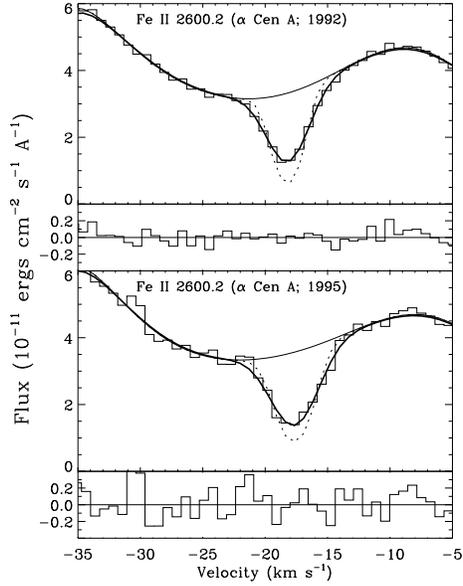}
\caption{Fits to the interstellar Fe~II $\lambda$2600.2 absorption line
  observed towards $\alpha$~Cen~A by HST/GHRS, once in 1992 and again in 1995.
  The dotted and thick solid lines are the fits before and after convolution
  with the instrumental profile, respectively \cite{jll96}.}
\end{figure}

     Figure~2 shows that the LISM absorption towards $\alpha$~Cen is
fitted nicely with only a single absorption component.  Many lines of sight
through the LISM have this desirable property, which simplifies the analysis
of the absorption line data.  However, many lines of sight also show multiple
components, including many short lines of sight to very nearby stars.
Figure~3 shows the Mg~II h \& k absorption lines observed towards 61~Cyg~A,
a star only 3.5~pc away \cite{bew98}.  Two LISM absorption components are
required to fit these two lines.  Other very nearby stars with LISM
absorption features indicating more than one component include Sirius
($d=2.6$~pc) \cite{rl94}, Procyon ($d=3.5$~pc) \cite{jll95}, and Altair
($d=5.1$~pc) \cite{rl95}.
\begin{figure}
\includegraphics[scale=0.5,trim=150 50 0 0]{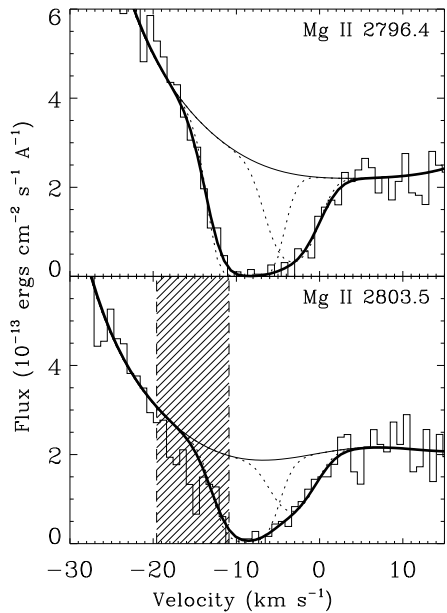}
\caption{Two-component fits to the interstellar Mg~II absorption lines
  seen towards 61~Cyg~A, with the individual components shown as dotted lines
  and the combination of the two components shown as thick solid lines
  (after convolution with the instrumental profile).  The shaded region was
  ignored in the fit due to contamination from a stellar absorption line
  \cite{bew98}.}
\end{figure}

     The existence of these multiple velocity components means that the LISM
is not characterized by a single velocity vector.  However, in all the
multiple-component cases mentioned above the components are not widely
separated, and this is generally the case within the LISM.  Thus, the
different velocity vectors are not in widely disparate directions or with
very different speeds.  Instead, they are all presumably minor variations
of basically the same vector.  Whether the individual components represent
different, physically distinct clouds is a complicated question that we will
return to in section 3.3.

     Besides their usefulness in establishing the velocity structure of
LISM lines of sight, the metal lines are also useful in measuring the
temperature ($T$) and nonthermal velocity ($\xi$) of the LISM.  Figure~4
shows how this is done for the case of the $\alpha$~Cen line of sight
\cite{jll96}.  Based on the measured Doppler parameters and equation (1),
$\xi$ is plotted versus $T$ for Fe~II, Mg~II, and D~I.  Thanks to the
different atomic weights of these elements, the curves have different
slopes.  The intersection of the curves indicates the $T$ and $\xi$
value for the line of sight.  Values of $T=5400\pm 500$~K and
$\xi=1.20\pm 0.25$ km~s$^{-1}$ are suggested by Figure~4.
\begin{figure}
\includegraphics[angle=90,scale=0.5,trim=40 0 0 -10]{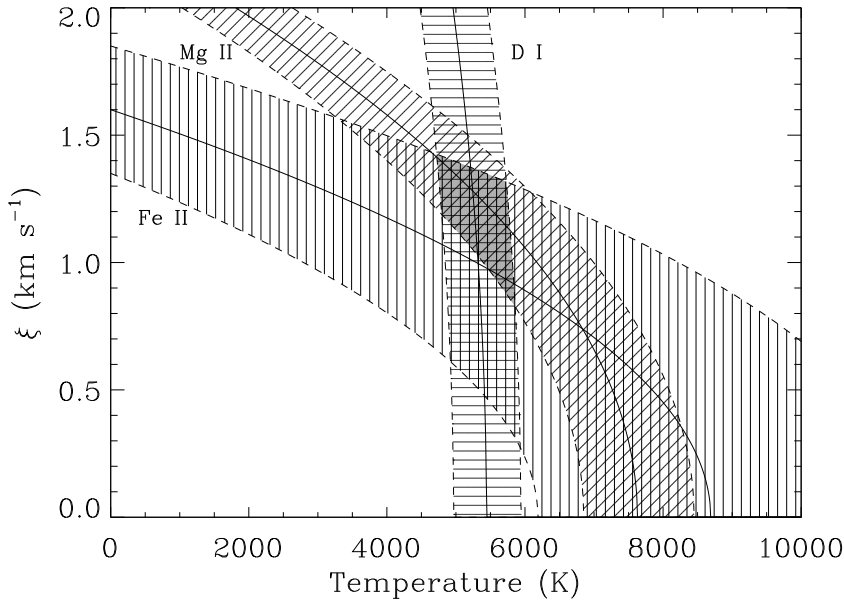}
\caption{Nonthermal velocities ($\xi$) are plotted versus temperature,
  based on the Doppler parameters and their uncertainties measured from
  absorption lines of Fe~II, Mg~II, and D~I observed towards $\alpha$~Cen.
  The shaded area where the three curves overlap indicates the actual
  temperature and $\xi$ value of the interstellar material \cite{jll96}.}
\end{figure}

\section{THE PHYSICAL STRUCTURE OF THE LISM}

\subsection{The Local Interstellar Cloud}

     The cloud within the LISM in which the Sun resides is generally
referred to as the Local Interstellar Cloud (LIC).  The properties of the
LIC have been studied in two fundamentally different ways.  One method is by
studying LISM absorption lines, as described in section 2.  The other method
relies on observations of interstellar gas within the heliosphere.
Detection of solar Ly$\alpha$ emission scattered back toward Earth by
interstellar H~I flowing through the heliosphere provided the first evidence
that the surrounding ISM was at least partially neutral \cite{jlb71}, and
Ly$\alpha$ backscatter within the heliosphere still remains a valuable source
of information about the LIC and its interaction with the solar wind.
The SWAN instrument on the SOHO satellite has provided the most recent
observations of this sort \cite{eq99,eq00}.  There are also direct
{\em in situ} studies of interstellar atoms within the heliosphere
that have been made by interplanetary spacecraft.  Of particular note are
the observations from the {\em Ulysses} satellite \cite{mw93,mw96}.

     Both measurements of ISM material flowing through the heliosphere and
LISM absorption line studies have been used to estimate the direction and
magnitude of the LIC vector \cite{mw93,mw96,rl95,eq00}, and the resulting LIC
vectors are in good agreement.  The vector derived from the absorption lines
has a magnitude of 25.7 km~s$^{-1}$ directed towards Galactic coordinates
$l=186.1^{\circ}$ and $b=-16.4^{\circ}$ \cite{rl95,rl92}.  With this vector,
LISM absorption from the LIC can be identified for any line of sight by its
centroid velocity.

     By measuring LIC column densities for many lines of sight in many
different directions, a three dimensional model of the shape of the LIC can
be constructed \cite{sr00}.  Figure~5 shows a map of the H I column density
between the Sun and the edge of the cloud in Galactic coordinates.  The
locations of various lines of sight used in the construction of the model
are indicated.  Observations from the {\em Extreme Ultraviolet Explorer}
(EUVE) and some optical Ca~II measurements were used to supplement the HST
data and fill in gaps left by the HST-observed lines of sight.  The largest
column density is about $2\times 10^{18}$ towards Galactic coordinates
$l=157^{\circ}$ and $b=-25^{\circ}$.  There is very little if any column in
the direction of the Galactic Center ($l=0^{\circ}$), indicating we are very
near the edge of the cloud in this direction.  This would be the direction of
the so-called G cloud, which will be discussed in section 3.2.
\begin{figure}
\includegraphics[scale=0.5,trim=40 110 0 120]{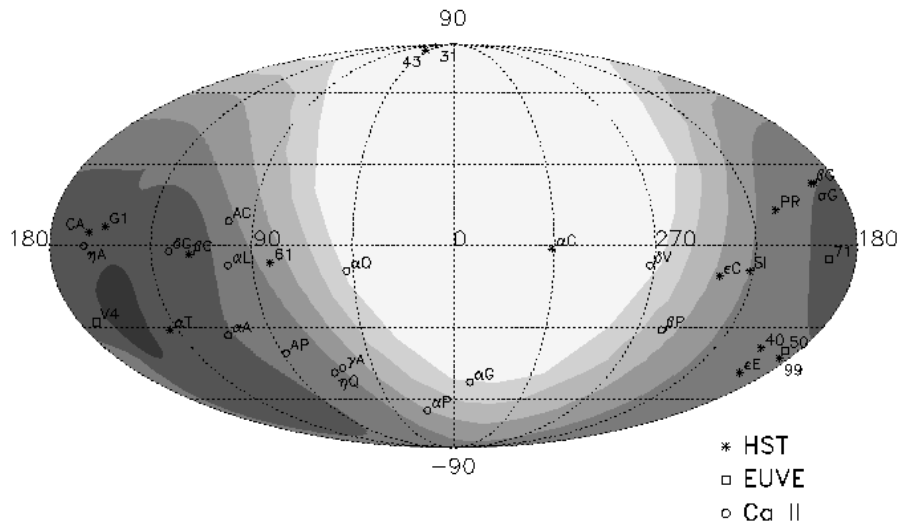}
\caption{Map in Galactic coordinates of the H~I column density from the Sun
  to the edge of the LIC, with the contour edges from darkest to lightest
  representing columns (in units of $10^{18}$ cm$^{-2}$) of 2.0, 1.0, 0.5,
  0.25, 0.10, and 0.05.  Positions of stars used in the creation of this
  3D LIC model are also plotted, with the symbol indicating if the data
  was obtained by HST, EUVE, or by ground-based observations of Ca~II
  \cite{sr00}.}
\end{figure}

     The highest average H~I densities seen towards the nearest stars are of
order $n({\rm H~I})=0.1$ cm$^{-3}$, so we assume this is representative of
the average density within the cloud.  Estimates of $n({\rm H~I})$ from
heliospheric observations suggest somewhat higher densities of
$n({\rm H~I})\approx 0.2$ cm$^{-3}$ \cite{eq94,vvi99a}, possibly suggesting
the presence of density and/or hydrogen ionization state variations within
the LIC.  In any case, assuming $n({\rm H~I})=0.1$ cm$^{-3}$ as an average
LIC density, the observed column densities in Figure~5 translate to a cloud
about 5--7 pc across depending on the direction, with a volume of 93~pc$^{3}$,
and a total mass of about 0.32~M$_{\odot}$ \cite{sr00}.

     Temperatures have been measured for the LIC for many lines of sight
using the technique shown in Figure~4.  These temperatures range from
$T=6900\pm 380$~K towards Procyon \cite{jll95} to $T=9700\pm 900$~K towards
$\beta$~Cas \cite{ard97}, so it is possible that there is some variation
within the LIC.  The weighted mean and standard deviation of all the
LIC measurements is $T=8000\pm 1000$~K \cite{ard97,np97,bew98}.  Turbulent
velocities are not as well determined but generally $\xi \leq 2$ km~s$^{-1}$.

     The most recent temperature determination from {\em Ulysses}
observations of interstellar He~I within the heliosphere, $T=5800\pm 700$~K
\cite{mw96}, is lower than the absorption line measurements.  This might be
further evidence for a temperature gradient within the LIC and within the
LISM in general.  Because of the Sun's location very near the edge of the
LIC, most of the absorption line studies measuring LIC properties are
in a generally anti-Galactic center direction ($90^{\circ}<l<270^{\circ}$),
as shown in Figure~5.  Lines of sight roughly towards the Galactic center
sample G cloud material rather than LIC material (see section 3.2).  The G
cloud is cooler than the LIC so the {\em Ulysses} measurement and absorption
line measurements of the G and LIC clouds together suggest a possible
temperature gradient in the LISM, with lower temperatures towards the
Galactic Center and higher temperatures in the opposite direction where
most of the LIC resides.

     The ionization state of the LIC is an issue that remains complicated
both observationally and theoretically.  Many attempts to measure the
electron density ($n_{e}$) in the LIC have been based on the observed
ionization states of magnesium or sodium, but these measurements tend to have
large uncertainties since the ionization state dependence on $n_{e}$ is also
highly temperature dependent \cite{pcf94,rl97}.  Furthermore, there is still
some question as to whether the LIC is actually in ionization equilibrium
at all \cite{chl96}, and if it is not then it will not be possible to
estimate $n_{e}$ from the ionization state of any element.  A final
complication is that the ionization state of the LIC can be expected to
be variable within the cloud, since the ionization state at a given location
will depend on how shielded that location is from various photoionization
sources by intervening LISM material \cite{fcb88,kpc90}.

     The best measurement of $n_{e}$ in the LIC comes from observations of
C~II lines towards Capella, which are shown in Figure~6 \cite{bew97}.  The
C~II $\lambda$1335 transition is from the ground state, which is where most
C~II ions in the LISM will reside.  The C~II $\lambda$1336 transition is from
an excited state that is much less populated than the ground state,
explaining why the $\lambda$1336 line is so much weaker than the
$\lambda$1335 line.  The excited state represented by the $\lambda$1336 line
is populated by collisions with electrons, and so the column density ratio of
the two C~II lines will depend on the electron density in the LIC.  The
advantage of this method for measuring $n_{e}$ is that it is not
temperature dependent and does not assume ionization equilibrium.
However, since the $\lambda$1335 line is saturated, the ground state C~II
column density has large uncertainties and therefore so does the derived
electron density:  $n_{e}=0.11^{+0.12}_{-0.06}$ cm$^{-3}$ \cite{bew97}.
This value is roughly consistent with estimates of the proton density that
have been made based on heliospheric observations \cite{vvi99a}.
\begin{figure}
\includegraphics[angle=90,scale=0.5,trim=40 0 0 -10]{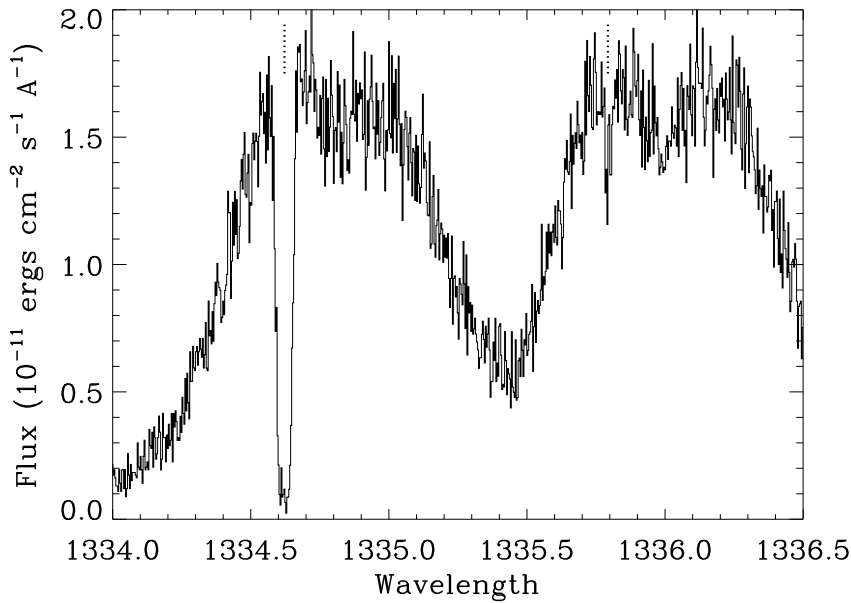}
\caption{HST/GHRS spectrum of the C~II $\lambda\lambda$1335, 1336 emission
  lines of Capella.  Interstellar C~II absorption features are marked with
  vertical dotted lines \cite{bew97}.}
\end{figure}

     The electron density in the LIC is comparable to the average LIC
$n({\rm H~I})$ value quoted above.  This means that hydrogen in the LIC is
roughly half-ionized since hydrogen will be the dominant source of electrons
in the LIC.  This is too high of an ionization state to be explained by
collisional ionization at a temperature of about 8000~K.  It has therefore
been generally assumed that the extreme UV background flux provided by
surrounding hot stars and hot ISM photoionizes the LISM and determines its
ionization state.  Observations from EUVE suggest that the photoionizing flux
from hot stars can indeed explain the half-ionized state of hydrogen
\cite{chl96}.

     However, EUVE has also found that helium is also roughly half-ionized in
the LISM, which is much harder to explain due to its higher ionization
potential \cite{jd95,jbh95,tl96}.  The stellar EUV background flux cannot
seem to explain the He ionization state, but while the stellar sources of
ionizing photons are known well, the diffuse EUV background provided by the
hot ISM in the local bubble and its interaction with neutral clouds is not
well known, so the issue of whether the ionization state of the local cloud
is entirely determined by photoionization is still open.  An alternative
theory is that the local cloud is not in ionization equilibrium at all, but
was instead collisionally ionized a million years or so ago by a passing
shock wave, perhaps from a nearby supernova, and still has not recombined to
an equilibrium state \cite{chl96}.

\subsection{The G Cloud}

     Figure~7 shows the outline of the LIC in the Galactic plane based on
the model in Figure~5 described above \cite{sr00}.  The Sun is very near
the edge of the cloud in the Galactic Center direction.  The placement of
the Sun at this location so near the edge is based on the discovery that
when you look in this direction you do not see material moving at the speed
of the local cloud, but instead you see material moving at a slightly
faster velocity \cite{rl92}.  Thus, it was proposed that there is a
different cloud that is observed in this direction, the so-called "G cloud",
and the Sun must be near the edge of its cloud to explain why absorption is
only observed from the G cloud in that direction.  The shape of the G cloud
is unknown, so the contour in Figure~7 is only a schematic representation
\cite{bew00b}.  The velocity vector derived for the G cloud is towards
$l=184.5^{\circ}$ and $b=-20.5^{\circ}$ with a magnitude of
29.4 km~s$^{-1}$ \cite{rl92}.  The LIC and G cloud vectors are illustrated
in Figure~7.  They are very similar, but the G cloud vector is slightly
faster:  29.4 km~s$^{-1}$ compared with 25.7 km~s$^{-1}$ for the LIC.
\begin{figure}
\includegraphics[scale=0.7,trim=30 30 0 0]{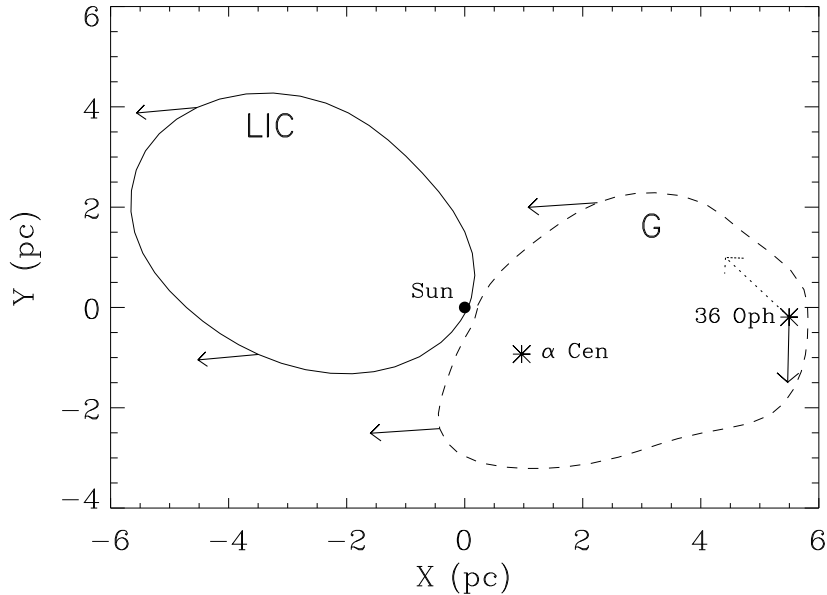}
\caption{A map showing the locations of the Sun, $\alpha$~Cen, 36~Oph, the
  LIC (solid line), and G cloud (dashed line), projected into the Galactic
  plane, where Galactic Center is to the right.  The LIC shape is estimated
  from a 3D model \cite{sr00}, while the G cloud shape is only a
  rough estimate.  The solid arrows indicate the velocity vectors of the LIC,
  the G cloud, and 36~Oph relative to the Sun, with the length of the arrow
  being proportional to the speed.  The dotted line indicates the
  G cloud vector in the rest frame of 36~Oph, thereby indicating the
  orientation of the star's astrosphere \cite{bew00b}.}
\end{figure}

     The evidence for the G cloud is best illustrated by Figure~8.  These
are HST/STIS observations of absorption seen towards 36 Oph, a star which
is only $12^{\circ}$ from the upwind direction of the LIC vector.
Since both {\em Ulysses} measurements and analyses of ISM absorption suggest
that the LIC moves with a speed of 26 km~s$^{-1}$, one would expect to see
absorption at about $-26$ km~s$^{-1}$ in this direction, or slightly less due
to projection effects.  But that is not what is observed.  Instead, focusing
on the Fe~II absorption, the absorption is centered at about
$-29$ km~s$^{-1}$, at the velocity predicted by the G cloud vector, and no
absorption component is observed centered on the expected LIC velocity.  The
D~I and Mg~II absorption is also centered at about $-29$ km~s$^{-1}$,
although for Mg~II this fact is not as apparent thanks to the steep gradient
of the background stellar emission.  In order to explain the lack of any LIC
absorption in Figure~8, the Sun must be within 0.19~pc of the edge of the
LIC, which the Sun will reach in less than 7400 years.  The faster G cloud
will be overtaking the LIC, possibly creating an interesting interaction
region between the two clouds \cite{sg96}.

\begin{figure}
\includegraphics[scale=0.5,trim=110 30 0 0]{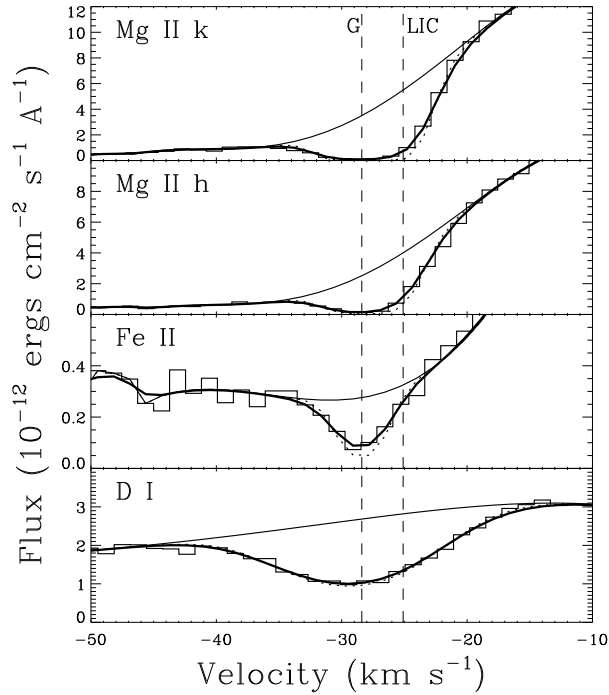}
\caption{Single component absorption line fits to the following lines
  observed towards 36~Oph:  Mg~II k $\lambda$2796.352,
  Mg~II h $\lambda$2803.531, Fe~II $\lambda$2600.173, and
  D~I $\lambda$1215.339.  The dotted and thick solid lines are before and
  after instrumental broadening, respectively.  Dashed lines show the
  projected velocities of the LIC and G cloud flow vectors for the 36~Oph
  line of sight \cite{bew00b}.}
\end{figure}

     Only two G cloud lines of sight have been studied in detail with HST,
the $\alpha$~Cen and 36~Oph lines of sight indicated in Figure~8.  These two
lines of sight suggest that the G cloud is cooler than the LIC, with an
average temperature of about $\sim 5650$~K compared with $\sim 8000$~K for
the LIC \cite{jll96,bew00b} (see, e.g., the measurement of $T$ for
$\alpha$~Cen in Fig.~4).  Abundances are also a bit different.  For example,
the Mg~II/H~I ratio is about four times higher in the G cloud than in the
LIC.  Considering the similarity of the LIC and G velocity vectors, it is
very reasonable to wonder if the two clouds are truly separate entities.
The fact that the physical properties of ISM material are a bit different in
the LIC and G cloud locations indicated in Figure~7 seems to suggest that
the LIC and G clouds really are physically distinct.  But there is more
recent evidence that seems to suggest the opposite, which will be discussed
in the following subsection.

\subsection{The Hyades Cloud}

     The G cloud is just to the right of the LIC in Figure~7.  It turns out
that there is another cloud that would be just to the left of the LIC in
this figure.  This cloud is called the Hyades cloud because its discovery
was based on HST/STIS observations of a bunch of stars in the Hyades cluster
about 50~pc away \cite{sr01}.  The positions of those stars are indicated
as circles in Figure~9.
\begin{figure}
\includegraphics[angle=90,scale=0.5,trim=70 0 70 -30]{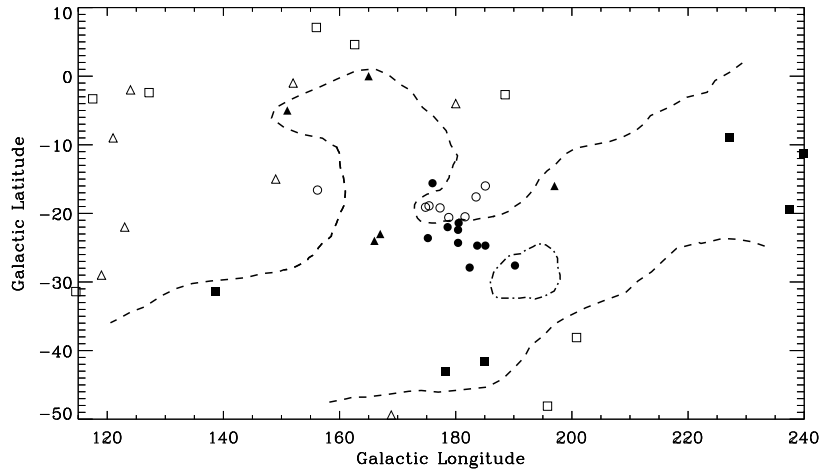}
\caption{Map in Galactic coordinates of Hyades stars observed by HST
  (circles), and other stars in the vicinity of the Hyades studied either
  from HST data (squares) or from ground-based
  observations of Ca~II (triangles).  Filled symbols indicate lines of sight
  with an absorption component consistent with the presence of Hyades cloud
  material.  The dashed lines represent an estimate for the outline of the
  cloud in the sky.  Another small cloud is seen towards one of the Hyades
  stars and its estimated outline is shown as a dot-dashed line \cite{sr01}.}
\end{figure}

     Figure~10 shows the Mg~II absorption lines observed toward 2 of the
Hyades stars \cite{sr01}.  Towards HD~29225 only one absorption component is
observed, which is associated with the LIC, while towards HD~29419 an
additional component is observed blueward of the LIC component by about
10 km~s$^{-1}$.  It is this second absorption component that is associated
with the Hyades cloud.  In Figure~9, filled circles indicate Hyades stars
that show the Hyades cloud absorption and open circles show only LIC
absorption.  Building on this work, other nearby lines of sight analyzed
previously were inspected in order to see which show absorption components
that might be from the Hyades cloud.  The squares in Figure~9 are other
HST-observed lines of sight, while the triangles are lines of sight for
which there are ground-based Ca~II data.  Filled symbols indicate lines
of sight with an absorption component just blueward of of the LIC component
that could be interpreted as being Hyades cloud absorption, and the dashed
line is an estimate of the apparent outline of the Hyades cloud on the sky,
which appears to have an extended, filamentary appearance.  One of the
Hyades stars (HD~28736) shows a third absorption component indicating a
third cloud, the location of which is indicated by the dot-dashed line
\cite{sr01}.
\begin{figure}
\includegraphics[angle=90,scale=0.45,trim=20 0 20 0]{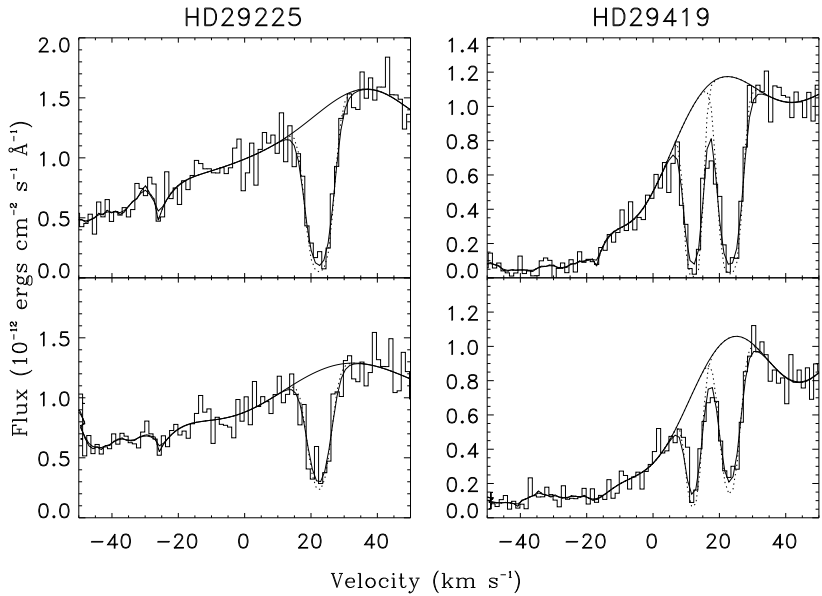}
\caption{Fits to the interstellar Mg~II h (bottom panels) and k (top panels)
  absorption lines observed towards two Hyades stars, one of which (HD~29225)
  shows only LIC absorption and one (HD~29419) which shows both LIC and
  Hyades absorption components.  Dotted and thick solid lines are the fits
  before and after correction for instrumental broadening, respectively
  \cite{sr01}.}
\end{figure}

     Although the distance to the Hyades stars is about 50~pc, there
is reason to believe the cloud is much closer.  For one thing, the Mg~II
column density through the cloud is less than that of the LIC so it seems
likely that the cloud is no larger than the LIC, but its apparent size in
Figure~9 is quite large, suggesting that it is nearby.  Also, a few of the
other stars with possible Hyades cloud absorption components are much closer
than the Hyades stars, including $\kappa$~Cet which is only 9.2~pc away.
The second absorption component observed towards Sirius could also possibly
be Hyades cloud absorption \cite{rl94}, and Sirius is only 2.6~pc away.
However, the further the angular distance from the Hyades, the more different
the projected velocity of the cloud will be, and the more uncertain the
identification with the Hyades cloud.

     The Hyades data is also useful for studying properties of the LIC.
In Figure~11, the positions of the Hyades stars are shown as dots.  The
grey scale shows the hydrogen column predicted by the LIC model in
Figure~4 \cite{sr00}.  The maximum LIC column happens to be in the field of
view at $l=157^{\circ}$ and $b=-25^{\circ}$.  The crude contours in
Figure~11 indicate the LIC Mg~II column density suggested by the Hyades
data.  The column density increases in the expected direction towards where
the LIC H~I column is predicted to be at its highest.  Another thing this data
set is useful for is to look for evidence of small scale structure within
the LIC.  If a lot of scatter was observed in the Mg~II columns of these
very nearby stars, that would indicate that the LIC has a lot of small scale
internal structure.  However, statistically significant scatter is not
detected for the very close star pairings.  The only clear variation of
Mg~II column within the Hyades data set is that of the general large scale
gradient shown in Figure~11 \cite{sr01}.
\begin{figure}
\includegraphics[angle=90,scale=0.5,trim=10 0 -10 -90]{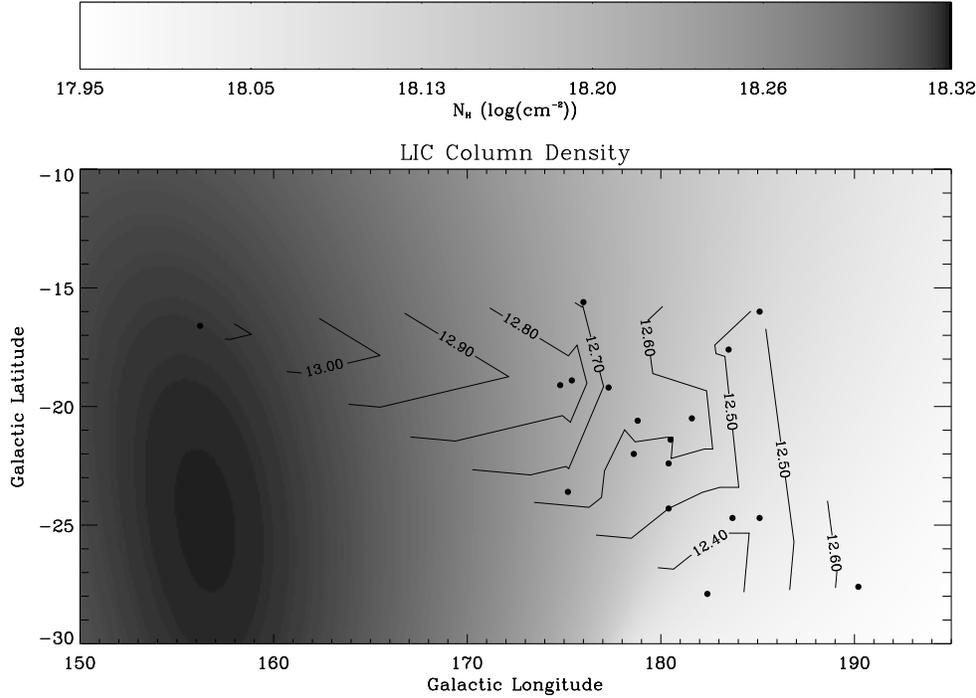}
\caption{Contours of the Mg~II column density of the LIC towards the Hyades
  stars, compared with the H~I column density predicted by a model of the
  LIC \cite{sr00}, which is indicated by shading.  The Mg~II gradient is
  consistent with the H~I gradient \cite{sr01}.}
\end{figure}

     The consistency of the Mg~II gradient with the expected LIC gradient
in Figure~11 is further support for that component actually being LIC.  This
additional support is useful because the centroid velocities of the LIC
components observed towards the Hyades are not exactly what the LIC vector
predicts they should be.  Figure~12 plots the observed velocities of the
LIC components seen towards the Hyades stars versus the velocities
predicted by the LIC vector.  The Hyades stars are almost directly in the
downwind direction relative to the LIC vector, so one would have expected
to see LIC velocities close to 26 km~s$^{-1}$, but the observed velocities
are too low by about 3 km~s$^{-1}$ on average.  This discrepancy mirrors the
discrepancy observed in the upwind direction towards 36~Oph, where the
observed velocity is 3 km~s$^{-1}$ too {\em high}.  The interpretation
for the upwind discrepancy was the existence of a distinct G cloud in
that direction.  However, the existence of a similar discrepancy downwind
but with the observed velocities being too low instead of too high suggests
that maybe we are detecting a velocity gradient within the LISM.  The
suggested gradient implies faster velocities upwind and slower velocities
downwind, meaning the LISM material is being compressed.  Perhaps this
general LISM velocity gradient extends even further to the Hyades cloud,
which is further downwind and does in fact appear to be slower than what we
have been calling the LIC and G clouds.
\begin{figure}
\includegraphics[scale=0.6,trim=40 30 0 0]{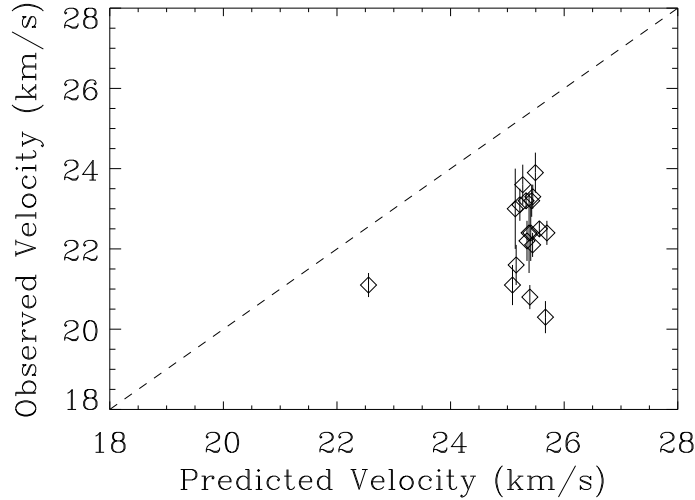}
\caption{Comparison of the observed velocity of the LIC component observed
  towards the Hyades stars with the value predicted by the LIC flow vector,
  showing that the observed velocities are too low by $\sim 3$ km~s$^{-1}$
  on average.}
\end{figure}

     The Hyades data appear to call into question the reality of a G cloud
that is a physically distinct entity from the LIC, and it also calls into
question the interpretation of other supposedly non-LIC LISM components;
the second component seen towards 61~Cyg in Figure~3, for example.  The
LIC component in Figure~3 is the one at $-9$ km~s$^{-1}$.  Is the second,
redshifted component really a separate cloud, or is there a velocity gradient
within the LIC that is responsible for the asymmetry in the Mg~II absorption
lines that leads to an erroneous two component interpretation?  A useful
exercise for the future will be to try to refine the LIC
vector to see if it can be altered to be consistent with the Hyades
velocities, while still preserving agreement with LIC velocities measured
for other lines of sight.  If this exercise is unsuccessful, it may be
necessary to abandon the notion of LISM clouds with uniform flow vectors,
which would greatly complicate interpretaions of LISM absorption line data.

\subsection{Unusual Lines of Sight}

     There are a few lines of sight observed by HST that suggest that there
is some nearby interstellar material with properties atypical for the LISM.
These observations further serve to illustrate the complexity of LISM
structure, and so we now briefly discuss these unusual lines of sight.

\subsubsection{HD 82558}

     With a measured H~I column of $\log N({\rm H~I})=19.05\pm 0.15$, the
18.3~pc line of sight to the K2~V star HD~82558 may have the highest
average density yet observed within the LISM,
$n({\rm H~I})\approx 0.2$ cm$^{-3}$ \cite{bew00a}.  Very large column
densities have also been observed towards the nearby ($d=14.3$~pc) A star
$\alpha$~Oph \cite{pcf87}, but much of this absorption may be circumstellar,
as many A stars are surrounded by circumstellar material \cite{hh99}.  The
$n({\rm H~I})\approx 0.2$ cm$^{-3}$ average density observed towards HD~82558
is only a factor of 2 higher than the average LIC density quoted in
section 3.1 ($n({\rm H~I})=0.1$ cm$^{-3}$), but this average towards HD~82558
is computed assuming that the H~I is evenly distributed along the entire
18.3~pc line of sight to HD~82558, which is very unlikely.  High column
densities have not been detected for any other nearby star in the general
direction of HD~82558, meaning that the cloud responsible for the absorption
is probably neither very large nor very close.  Thus, the H~I encountered
towards HD~82558 that is producing so much absorption probably does not
occupy a large fraction of the line of sight, and the actual density of the
H~I is probably more like $n({\rm H~I})\geq 0.5$ cm$^{-3}$, significantly
higher than average densities within the LIC or G clouds.  What makes this
high density line of sight even more noteworthy is that HD~82558 is only
$39^{\circ}$ from the ``interstellar tunnel'' towards $\epsilon$~CMa
\cite{cg95}, where column densities are an order of magnitude lower towards
stars an order of magnitude farther away \cite{bew00a}!

\subsubsection{$\beta$ Ceti}

     The $\beta$~Ceti line of sight is remarkable for having a very high
magnesium abundance \cite{np97}.  Metal abundances within the LISM are
generally found to be lower than solar, presumably due to depletion of metals
onto dust grains \cite{pcf99}.  A crude Mg abundance can be defined for a
line of sight as ${\rm A(Mg)}\equiv N({\rm Mg~II})/N({\rm H~I})$.  This
assumes that Mg~II is the dominant ionization state of Mg (probably a good
assumption) and H~I is the dominant ionization state of H (probably only good
to within a factor of 2 --- see section 3.1).  A logarithmic depletion
quantity can then be defined as
${\rm D(Mg)}\equiv \log {\rm A(Mg)} - \log {\rm A(Mg)_{\odot}}$, where
${\rm A(Mg)_{\odot}}=-4.41$ is the solar Mg abundance \cite{ea89}.
Typical values for the LIC are ${\rm D(Mg)}\approx -1.2$ \cite{np97,bew98},
meaning Mg in the LIC is underabundant by a factor of 15 due to depletion
onto dust grains.  Depletions are lower in the G cloud (see section 3.2),
with ${\rm D(Mg)}\approx -0.6$ \cite{jll96,bew00b}.

     However, the 29~pc line of sight to the K0~III star $\beta$~Cet has a
remarkable depletion value of ${\rm D(Mg)}= +0.30\pm 0.15$, suggesting that
if anything Mg is overabundant rather than depleted \cite{np97}.  Significant
ionization of H could at least remove the appearance of overabundance, but
it seems clear that Mg is not significantly depleted towards $\beta$~Cet.
The reason for this remains a mystery.  We can only speculate that perhaps
shocks in that direction have dissociated the dust grains and released all
the previously depleted Mg back into the gas phase.  Finally, we note that at
least 2 and perhaps 3 velocity components are observed towards $\beta$~Cet,
with the dominant component being at the velocity predicted by the LIC
vector.  However, this component's identification with the LIC is
questionable considering the wildly discrepant D(Mg) value.

\subsubsection{70~Ophiuchi}

     Figure~13 shows the Mg~II absorption observed towards 70~Oph, a K0~V
star only 5.1~pc away that is close to the previously discussed 36~Oph line
of sight in both distance and
direction (see Fig.~7).  Like 36~Oph, 70~Oph shows strong absorption at the
position predicted by the G cloud vector, although the column toward 70 Oph
is significantly higher.  However, for 70~Oph there are two additional
absorption components blueward of the saturated G cloud component.  It is not
unheard of to see multiple velocity components towards stars even as close
as 70~Oph, but the velocity components are generally separated by less than
10 km~s$^{-1}$.  However, for the most blueshifted absorption component
towards 70~Oph, the separation from the main component is about
18 km~s$^{-1}$.  This suggests that at least for this one line of sight
there is apparently some LISM material with a velocity vector significantly
different from the LIC and G cloud vectors.  The origin of this anomalously
fast LISM absorption component is unknown.
\begin{figure}
\includegraphics[scale=0.4,trim=20 40 0 0]{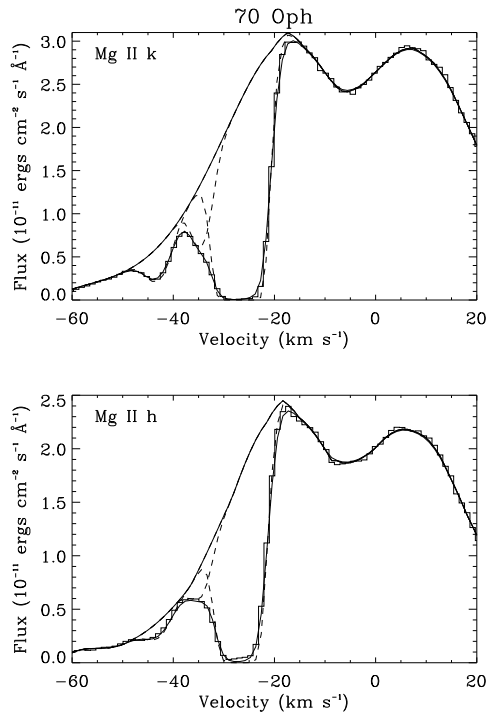}
\caption{Three-component fits to the LISM Mg~II h \& k absorption lines
  observed towards 70~Oph, with the individual components shown as dotted
  lines and the combination of the two components shown as thick solid lines
  (after convolution with the instrumental profile).}
\end{figure}

\section{HELIOSPHERIC AND ASTROSPHERIC ABSORPTION}

     One of the primary applications of LISM studies is to better understand
how the LISM interacts with the winds of the stars contained within it,
including our own Sun.  It turns out that the same HST H~I Ly$\alpha$ data
that is used to study the LISM can also be used to study both the
heliospheric interaction region surrounding our Sun and analogous
``astrospheric'' interaction regions around other nearby stars.  Figure~14
shows the H~I density and temperature within the heliosphere based on a
hydrodynamic model \cite{gpz96,bew00c}.  The heliopause (i.e., the contact
surface separating the solar wind and ISM plasma flows) and bow shock are
both visible at about 120~AU and 230~AU, respectively, in the upwind
direction.  Interstellar H~I crossing the bow shock is heated by charge
exchange processes \cite{gpz96,vbb93,vbb95}.  This heated heliospheric
hydrogen produces a substantial amount of Lyman-alpha absorption, enough to
be detectable in HST data despite being highly blended with the LISM
absorption.  Analogous material around other solar-like stars has also been
detected.
\begin{figure}
\includegraphics[scale=0.4,angle=-90,]{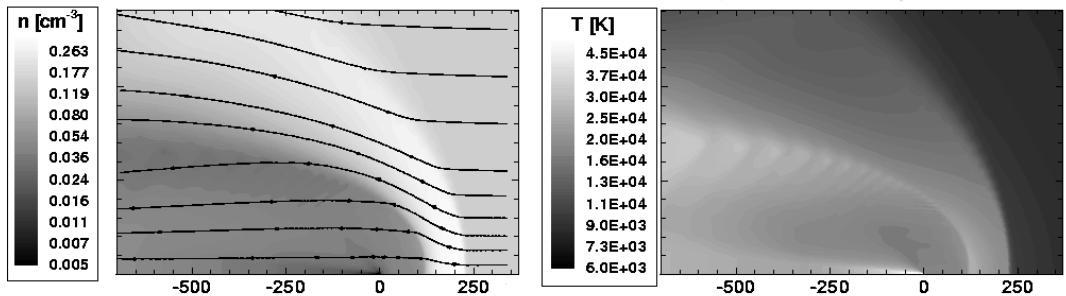}
\caption{The H~I density (left) and temperature (right) within the
  heliosphere predicted by a hydrodynamic model of the ISM--solar wind
  interaction, with streamlines shown in the left panel.  The heliopause
  and bow shock are both visible at about 120~AU and 230~AU, respectively,
  in the upwind direction.}
\end{figure}

     The four panels in the middle of Figure~15 show schematically what
happens to a stellar Ly$\alpha$ emission profile as it travels from the star
towards the Sun.  (This particular example is modeled on the $\alpha$~Cen
line of sight.)  The Ly$\alpha$ profile first has to traverse the hot
hydrogen in the star's astrosphere which erases the central part of the line.
The Ly$\alpha$ emission then makes the long interstellar journey, resulting
in additional absorption, including some D~I absorption.  Finally, the
profile has to travel through the hot hydrogen in the heliosphere, which
results in additional absorption on the red side of the line.  The reason the
heliospheric absorption is redshifted is because of the deceleration of
interstellar material as it crosses the bow shock.  Conversely, astrospheric
absorption will be blueshifted relative to the ISM absorption since we are
viewing that absorption from outside the astrosphere rather than inside. 
\begin{figure}
\includegraphics[scale=0.8,trim=160 70 0 40]{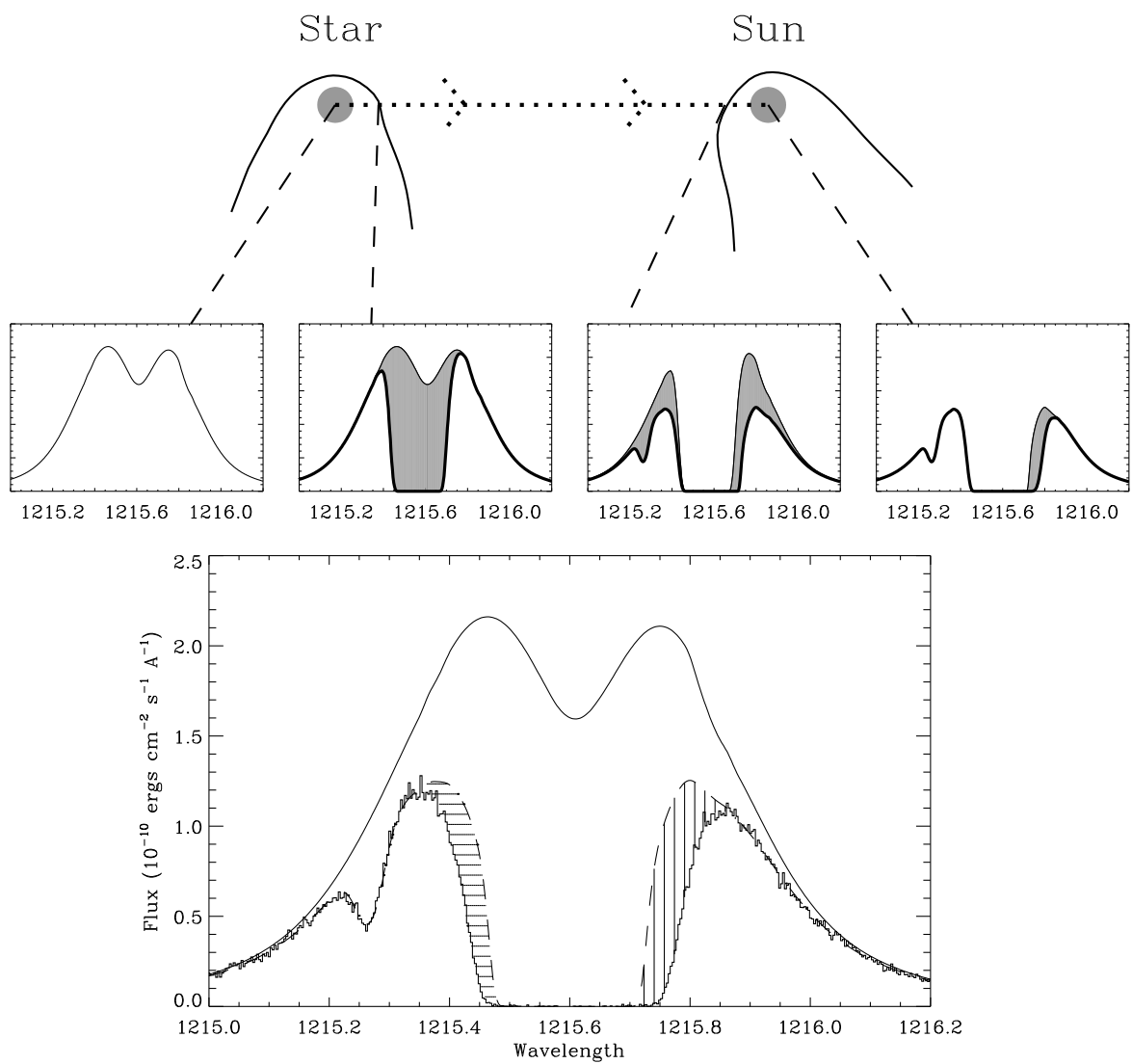}
\caption{Schematic diagram showing how a stellar Ly$\alpha$ profile changes
  from its initial appearance at the star and then through various regions
  that absorb parts of the profile before it reaches an observer at Earth:
  the stellar astrosphere, the LISM, and finally the heliosphere.  The lower
  panel shows the actual observed Ly$\alpha$ profile of $\alpha$~Cen~B.  The
  upper solid line is the assumed stellar emission profile and the dashed
  line is the ISM absorption alone.  The excess absorption is due to
  heliospheric H~I (vertical lines) and astrospheric H~I (horizontal lines).}
\end{figure}

     The bottom panel of Figure~15 shows the observed Ly$\alpha$ profile
of $\alpha$~Cen~B \cite{jll96}.  The upper solid line is the assumed stellar
line profile.  The dashed line is the LISM absorption alone, which is
determined by forcing the interstellar H~I absorption to have a Doppler
parameter and central velocity consistent with D~I and other LISM lines (see
section 2.1).  There is excess absorption on both sides of the LISM
absorption, which cannot be accounted for by the single LISM component
(G cloud) observed towards $\alpha$~Cen.  The red side excess is best
interpreted as heliospheric absorption and the blue side excess is best
interpreted as astrospheric absorption \cite{jll96,kgg97}.

     Heliospheric absorption has been detected towards three stars:
$\alpha$~Cen \cite{jll96}, Sirius \cite{vvi99b}, and 36~Oph \cite{bew00b}.
Comparisons with heliospheric models have shown that the observed
heliospheric H~I absorption is consistent with the predictions of the
hydrodynamic models, and can potentially be used as a diagnostic for
various input parameters for the models \cite{bew00b,kgg97,vvi99b},
such as the uncertain pressure provided by cosmic rays and the interstellar
magnetic field.  Astrospheric absorption has by now been detected towards
seven stars, which are listed in Table~1, although the $\lambda$~And and
40~Eri~A detections should be regarded as tentative \cite{bew98,bew96}.
These astrospheric detections represent the first detections of winds
analogous to the solar wind from solar-like main sequence stars.
\begin{table}[t]
\caption{Compilation of Astrospheric Detections}
\begin{tabular}{lccccc} \hline
Star & Spectral & $d$ & $V_{ISM}$ & $\theta$ & Reference \\
 & Type & (pc) & (km~s$^{-1}$) & (deg) & \\
\hline
$\alpha$ Cen  & G2 V+K0 V   & 1.3 & 25 & 79 & \cite{jll96} \\
$\epsilon$ Eri& K1 V        & 3.2 & 27 & 76 & \cite{ard97} \\
61 Cyg        & K5 V+K7 V   & 3.5 & 86 & 46 & \cite{bew98} \\
$\epsilon$ Ind& K5 V        & 3.6 & 68 & 64 & \cite{bew96} \\
40 Eri A      & K1 V        & 5.0 &127 & 59 & \cite{bew98} \\
36 Oph        & K1 V+K1 V   & 5.5 & 40 &134 & \cite{bew00b} \\
$\lambda$ And & G8 IV-III+? & 26  & 53 & 89 & \cite{bew96}   \\
\hline
\end{tabular}
\end{table}

     The wind speed ($V_{ISM}$) and orientation angle relative to the
Sun-star line of sight ($\theta$) of the LISM wind vector seen by each star
can be computed from our knowledge of the LIC and/or G cloud flow vectors and
the known proper motion and radial velocity of these stars.  An example of
this is illustrated in Figure~7 for the case of 36~Oph.  The $V_{ISM}$ and
$\theta$ values for each star with detected astrospheric absorption are
listed in Table~1.  The astrospheric observations show that the temperature
of the astrospheric material is clearly higher for stars with faster $V_{ISM}$
values, consistent with theoretical expectations since a higher ISM wind
speed will result in more shock heating at the bow shock \cite{bew98}.

     With knowledge of $V_{ISM}$ and $\theta$ for a star, one can compute
hydrodynamic models of the astrosphere analogous to the heliospheric model
shown in Figure~14, and estimate predicted Ly$\alpha$ absorption from
these models.  The amount of absorption will depend on the mass loss rate
assumed for the star.  The larger the mass loss rate, the larger the
astrosphere, and the higher the astrospheric H~I column density.  Thus,
the astrospheric absorption can be used to obtain the first estimates of
mass-loss rates for nearby solar-like stars \cite{bew01,hrm01a,hrm01b}.

     Figure~16 reproduces the $\alpha$~Cen~B spectrum from Figure~15
and compares the observed astrospheric absorption with the predictions of
four models assuming four different mass-loss rates.  The model assuming
twice the solar mass-loss rate best fits the data.  Since $\alpha$~Cen is
actually a binary star consisting of two very solar-like stars with a
combined surface area about twice that of the Sun, this result is very
reasonable \cite{bew01}.
\begin{figure}
\includegraphics[scale=0.65,trim=30 40 0 20]{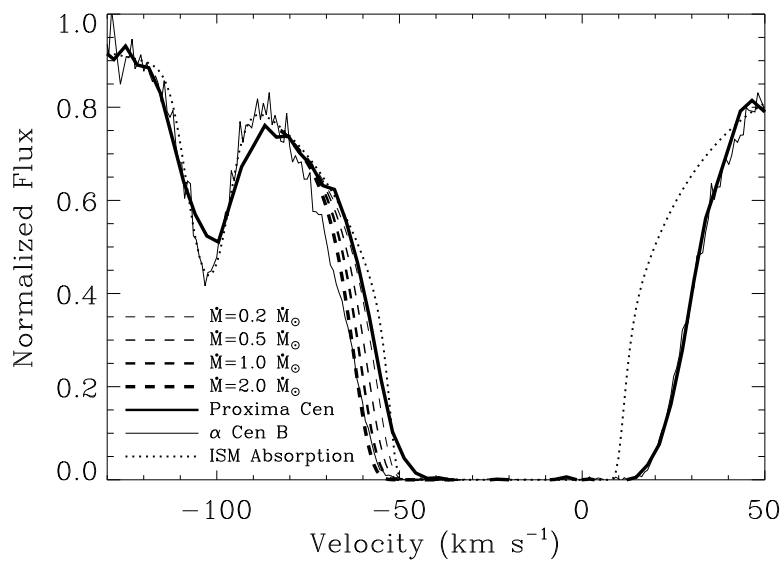}
\caption{The $\alpha$~Cen~B spectrum (thin solid line) and inferred ISM
  absorption (dotted line) are compared with a lower resolution STIS
  spectrum of $\alpha$~Cen's distant companion Proxima~Cen (thick solid
  line).  The dashed lines show the blue-side excess
  Ly$\alpha$ absorption predicted by four models of the Alpha/Proxima~Cen
  astrospheres, assuming four different mass loss rates.  The
  $2.0~\dot{M}_{\odot}$ model fits the $\alpha$~Cen spectrum reasonably
  well, and the $0.2~\dot{M}_{\odot}$ model represents an upper limit for
  the mass loss rate of Proxima~Cen \cite{bew01}.}
\end{figure}

     Also shown in Figure~16 is the Ly$\alpha$ line observed towards
Proxima~Cen, a distant companion to the two $\alpha$~Cen stars that will
not lie within the $\alpha$~Cen astrosphere.  The LISM absorption should
be the same, however, and the nearly identical amounts of D~I absorption
observed towards $\alpha$~Cen and Proxima~Cen suggests that this is the
case.  (The Proxima~Cen data have somewhat lower spectral resolution, which
explains why the the Proxima~Cen D~I profile is somewhat broader and
shallower.)  The blue-side excess absorption observed towards $\alpha$~Cen
that is being interpreted as astrospheric absorption is {\em not} observed
towards Proxima~Cen, which conclusively demonstrates that the excess
absorption towards $\alpha$~Cen must in fact be circumstellar.  If the
excess was somehow due to LISM or heliospheric material it should have
been observed towards both stars.

     There is no detected astrospheric absorption towards Proxima~Cen, and
the model predictions shown in Figure~16 suggest an upper limit for the
Proxima~Cen mass loss rate of a fifth the solar value \cite{bew01}.  A low
mass loss rate for Proxima~Cen is not surprising in that Proxima~Cen is a
tiny, dim star with a surface area 40 times lower than that of the Sun.
However, Proxima~Cen does have a surprisingly active corona, with frequent
large flares, so it was not obvious {\em a priori} that it would have a weak
wind.

     In principle, mass-loss rates can be estimated for all the stars in
Table~1.  A mass-loss rate of 0.5 solar has been estimated for $\epsilon$~Ind
and a rate of 5 times solar has been estimated for $\lambda$~And
\cite{hrm01a,hrm01b}.  Estimates for the other stars will surely follow.
Such studies, if expanded to include a sufficient number of stars,
could allow us for the first time to see how mass-loss varies with
stellar activity, age, spectral type, etc.  As a final comment, it is
remarkable that high resolution Ly$\alpha$ spectra of nearby stars are
relevant to studies of such a wide variety of topics:  stellar chromospheres,
D/H and its implications for cosmology and Galactic chemical evolution,
the structure and properties of the LISM in general, heliospheric physics,
and, as just discussed, stellar winds which are otherwise undetectable.

\end{document}